\documentclass[aps,prb,amsmath,twocolumn]{revtex4}

\usepackage{graphicx}
\usepackage{bm}

\begin{document}

\title{Crossed Andreev reflection and charge imbalance in diffusive NSN structures}

\author{Dmitri S. Golubev$^1$, Mikhail S. Kalenkov$^2$, and Andrei D.
Zaikin$^{1,2}$}
\affiliation{$^1$Forschungszentrum Karlsruhe,
Institut f\"ur Nanotechnologie, 76021 Karlsruhe}
\affiliation{$^2$I.E. Tamm Department of Theoretical Physics, P.N.
Lebedev Physics Institute, 119991 Moscow, Russia}

\begin{abstract}

We formulate a microscopic theory of non-local electron transport
in three-terminal diffusive normal-superconducting-normal (NSN)
structures with arbitrary interface transmissions. At low energies
$\varepsilon$ we predict strong enhancement of non-local spectral
conductance $g_{12} \propto 1/\varepsilon$ due to quantum
interference of electrons in disordered N-terminals. In contrast,
non-local resistance $R_{12}$ remains smooth at small
$\varepsilon$ and, furthermore, is found to depend neither on
parameters of NS interfaces nor on those of N-terminals. At higher
temperatures $R_{12}$ exhibits a peak caused by the trade-off
between charge imbalance and Andreev reflection. Our results are
in a good agreement with recent experimental observations and can
be used for quantitative analysis of future experiments.
\end{abstract}


\maketitle

In hybrid NS structures quasiparticle current flowing in a normal
metal is inevitably converted into that of Cooper pairs inside a
superconductor. For quasiparticle energies above the
superconducting gap $\varepsilon > \Delta$ this conversion is
accompanied by electron-hole (or charge) imbalance
\cite{CT} which relaxes inside a superconductor at a typical
inelastic length usually denoted as $\Lambda_{Q^*}$. As a result,
at temperatures near the critical one 
$T_C$ an electric field penetrates into a superconductor 
causing resistance enhancement for NS structures under consideration.

At subgap energies $\varepsilon < \Delta$ the physical picture
becomes entirely different. In this case
quasiparticle-to-Cooper-pair current conversion is provided by the
mechanism of Andreev reflection \cite{And}: A quasiparticle enters
the superconductor from the normal metal at a length of order of
the superconducting coherence length $\xi_S$, forms a Cooper pair
together with another quasiparticle, while a hole goes back into
the normal metal. Due to this process subgap conductance of the NS
structure remains non-zero down to $T=0$ \cite{BTK}. Furthermore,
in the presence of disorder this subgap conductance can be greatly
enhanced at low energies due to quantum interference effects
\cite{VZK,HN,Z}.

Further interesting effects may occur in three-terminal NSN
structures. Provided the distance between two N-terminals is
smaller than or comparable with $\xi_S$, electrons penetrating
into the superconductor from the first N-terminal may form Cooper
pairs with electrons from the second N-terminal. Then a hole goes
into the {\it second} N-metal making the charge transfer
effectively non-local. This important phenomenon of non-local (or
crossed) Andreev reflection (CAR) \cite{BF} enables direct
experimental realization of entanglement between electrons from
spatially separated N-terminals.

CAR was detected and investigated in several recent experiments
\cite{Beckmann,Teun,Venkat,Basel} by measuring the non-local
resistance of multiterminal NSN systems. The authors observed a
rich structure of different features many of which are still
waiting for their theoretical interpretation. Note that not only
CAR but also other physical processes contribute to the non-local
conductance $g_{12}$ thus making this interpretation rather
complicated. For instance, the contribution of elastic cotunneling
(EC) to $g_{12}$ exactly cancels that of CAR in the lowest order
in NS interface transmissions and at subgap energies \cite{FFH}.
This cancellation is lifted either in higher orders in barrier
transmissions \cite{KZ06} or in the presence of interactions,
e.g., with an effective external environment \cite{LY}, or under
external ac bias \cite{GZ09}.

Another important issue is the effect of disorder in metallic
terminals which needs to be analyzed for adequate interpretation
of experimental results \cite{Beckmann,Teun,Venkat,Basel}.
Although CAR in disordered NSN structures was already addressed in
a number of theoretical works \cite{Melin,Duhot,BG,Belzig,GZ07} in
various physical limits, we believe that general analysis of this
issue is still missing in the literature. For instance, the role
of disorder-induced electron interference \cite{VZK,HN,Z} in the
non-local subgap transport, the effect of high NS barrier
transmissions on CAR as well as some other features remain
unclear. Yet another important unresolved problem is to describe
an interplay between CAR and non-local charge imbalance. It was
demonstrated both experimentally \cite{Beckmann,Venkat} and
theoretically \cite{GZ07,KZ08} that such interplay may result in a
large non-local resistance peak which occurs at temperatures
slightly below the critical one. It was conjectured \cite{Venkat}
that the behavior of this peak is controlled by the charge
imbalance length $\Lambda_{Q^*}$ parametrically exceeding the
length scale $\sim \xi_S$ relevant for CAR. In this work we
develop a general theory of non-local electron transport in
diffusive NSN structures which enables one to clarify the above
issues and to formulate predictions to be tested in future
experiments.

\begin{figure}
\includegraphics[width=6.6cm]{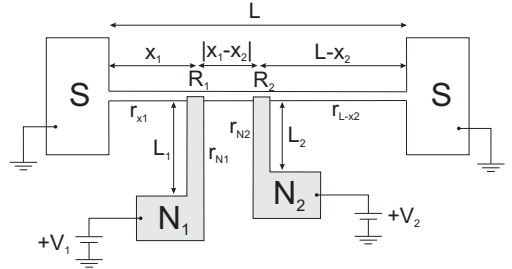}
\caption{NSN structure under consideration.}
\end{figure}

{\it Model and basic equations.} In what follows we will analyze a
multiterminal diffusive NSN structure schematically shown in Fig.
1. Two normal terminals $N_1$ and $N_2$ with resistances $r_{N_1}$
and $r_{N_2}$ and electric potentials $V_1$ and $V_2$ are
connected to a superconducting electrode of length $L$ with normal
state (Drude) resistance $r_L$ and electric potential $V=0$ via
small NS barriers with resistances $R_1$ and $R_2$ which can be
expressed via channel transmissions $T_{1,n}$ and $T_{2,n}$ of
these barriers as $1/R_{1(2)}=(e^2/\pi )\sum_n T_{1(2),n}$. For
the sake of definiteness in Fig. 1 we chose specific geometry
directly related, e.g., to experiments \cite{Venkat} where the
superconductor was fabricated in the form of a rather thin strip.
NS barriers are located at the points $\bm{r}_{1,2}=(x_{1,2},0,0)$
and the corresponding segments of a superconducting strip have
normal state resistances $r_{x_1}$ and $r_{L-x_2}$, see Fig. 1.

Our analysis is based on the quasiclassical Usadel equations for
the Green-Keldysh matrix functions $\check G$. In the absence of
interactions these equations read \cite{BWBSZ}
\begin{equation}
iD\nabla(\check G \nabla \check G) = \left[ \check\Sigma , \check
G \right], \quad \check G^2=1, \label{Usadel}
\end{equation}
where $[\check a, \check b]= \check a \check b - \check b \check
a$, $D$ is the diffusion constant and
\begin{equation}
\check G = \left(\begin{array}{cc}
\hat G^R & \hat G^K \\
0 & \hat G^A \\
\end{array}\right), \quad
\check \Sigma = \check 1 \left(\begin{array}{cc}
\varepsilon + eV & \Delta \\
-\Delta^* & -\varepsilon + eV\\
\end{array}\right)
\end{equation}
 are $4\times 4$ matrices in
Keldysh$\otimes$Nambu space, $\varepsilon$ is the quasiparticle
energy, $\Delta (T)$ is the superconducting order parameter which
will be considered real further below and $V$ is the electric
potential.

Far from the interfaces between metals the quasiclassical Green
functions $\check G$ coincide with their bulk equilibrium values.
Deep in the superconductor they read
\begin{equation}
\hat G^{R,A}_{S}= \pm\frac{\hat\tau_3\varepsilon
+i\hat\tau_2\Delta}{\sqrt{(\varepsilon\pm i \delta)^2 -
\Delta^2}}, \, \hat G^{K}_{S}=(\hat G^{R}_{S} - \hat G^{A}_{S}
)n(\varepsilon), \label{K}
\end{equation}
where $n(\varepsilon)=\tanh(\varepsilon/2T)$ and $\hat \tau_i$ are
Pauli matrices. In the normal terminals far from the tunnel
barriers one has
\begin{eqnarray} && \hat
G^{K}_{1,2}=2\left(
\begin{array}{cc} \tanh\frac{\varepsilon+eV_{1,2}}{2T} & 0 \\
0 & -\tanh\frac{\varepsilon-eV_{1,2}}{2T} \end{array}\right),
\end{eqnarray}
while the retarded and advanced Green functions  $\hat
G^{R,A}_{1,2}$ are set by the first Eq. (\ref{K}) with $\Delta
=0$. In the vicinity of the barriers the Green functions deviate
from the above equilibrium values and should be determined from
Eqs. (\ref{Usadel}) supplemented by appropriate boundary
conditions describing electron transfer across metallic
interfaces. For diffusive superconductors one finds
\cite{Nazarov99}
\begin{eqnarray}
\mathcal{A}_1 \sigma_1 \check G_1 \partial_x \check G_1 =
\mathcal{A}_1 \sigma_S \check G_S \partial_x \check G_S
\nonumber\\
= \frac{e^2}{\pi} \sum_{n} \frac{2T_{1,n} [\check G_1 ,
\check G_S]}{4+T_{1,n} (\{\check G_1 , \check G_S\} -2)}
\label{Nazarov}
\end{eqnarray}
for the first interface and similarly for the second one. Here
$\mathcal{A}_{1,2}$ are the barrier cross sections and
$\sigma_{S,1,2}$ are Drude conductivities of S- and N-terminals.

Having derived the Green-Keldysh functions $\check G$ one can
easily evaluate the current density $\bm{j}$ in our system with
the aid of the standard relation
\begin{equation}
\bm{j}= -\frac{\sigma}{8e} \int {\rm tr} [\hat \tau_3 (\check G
\nabla \check G)^K]d \varepsilon . \label{current}
\end{equation}

{\it Non-local spectral conductance.} The above general formalism
enables one to describe electron transport at arbitrary barrier
transmissions $T_{1,n}$ and $T_{2,n}$. Here we only assume that
both NS barriers are sufficiently small to provide $R_{1,2}\gg
r=\max(r_L,r_{N_1},r_{N_2})$. This condition allows to effectively
linearize Eqs. (\ref{Usadel}) and express the solution of
linearized Usadel equations via the diffuson ${\cal
D}^{\bm{r}\bm{r}'}(\omega)$ and the Cooperon ${\cal
C}^{\bm{r}\bm{r}'}(\omega)$. The diffuson satisfies the following
diffusion equation
\begin{eqnarray}
\left(-i\omega+\frac{1}{\tau_{Q^*}}-D\nabla^2\right)
{\cal D}^{\bm{r}\bm{r}'}(\omega)&=&\delta(\bm{r}-\bm{r}'),
\label{7}
\end{eqnarray}
while Cooperon is the solution of Eq. (\ref{7}) with effective
charge imbalance relaxation time $\tau_{Q^*}$ replaced by
dephasing time $\tau_\varphi$. At $T \sim T_C$  $\tau_{Q^*}$ depends on
the electron inelastic relaxation time $\tau_{in}$ as \cite{CT}
$\tau_{Q^*} \sim \tau_{in}T/\Delta (T)$.

Let us employ the standard representation of the Keldysh function
$\hat G^K=\hat G^R\hat h-\hat h\hat G^A$ with $\hat h=f_L\hat
1+f_T\hat\tau_3$, where $f_L$ and $f_T$ are respectively symmetric
and antisymmetric in energy parts of the distribution function.
Combining the above expression for $\hat G^K$ with Eq.
(\ref{current}) we define the current across the first barrier
\begin{eqnarray}
I_1=\frac{1}{2e}\int d\varepsilon
g_{1}(\varepsilon)[f_T^{N_1}(\varepsilon,\bm{r}_1)-f_T^S(\varepsilon,\bm{r}_1)].
\label{I1}
\end{eqnarray}
The spectral conductance $g_1(\varepsilon)$ is expressed via the
functions $\hat G^R,\hat G^A$. Solving Eq. (\ref{Usadel}) for
$\hat G^R$ and keeping terms up to the first order in $r/R_{1,2}$,
we find
\begin{eqnarray}
g_1(\varepsilon) &=& g^{\rm BTK}_1(\varepsilon)
+\frac{\theta(\Delta-|\varepsilon|)\Delta^2}{\Omega^2} \frac{{\rm
Re}\,{\cal
C}_1^{\bm{r}_1\bm{r}_1}(2\varepsilon)}{2e^2N_1R_1^2} \nonumber\\
&& +\,\frac{\Delta^2}{\Omega^2}\sum_{j=1,2}
 \frac{{\rm Re}\,{\cal C}_S^{\bm{r}_1\bm{r}_j}\big(2W(\varepsilon)\big)}{2e^2N_SR_j},
\label{g1}
\end{eqnarray}
where $W(\varepsilon)=i\Omega =i\sqrt{\Delta^2-\varepsilon^2}$ for
$|\varepsilon | <\Delta$, $W(\varepsilon )= |\Omega |\,{\rm
sign}\,\varepsilon$ for $|\varepsilon|>\Delta$, and $g_1^{\rm
BTK}(\varepsilon)$ is defined by the standard expression
\cite{BTK}
\begin{eqnarray}
g_1^{\rm BTK}(\varepsilon)=\frac{e^2}{\pi}\sum_n \left[
\frac{2T_{1,n}^2\theta(\Delta-| \varepsilon |)\Delta^2}{
T_{1,n}^2\varepsilon ^2+\left(2-T_{1,n}\right)^2\Omega^2} \right.
\nonumber\\
\left. +\,\frac{2T_{1,n}\theta(| \varepsilon |-\Delta)|
\varepsilon |}{T_{1,n}|\varepsilon|+\left(2-T_{1,n}\right)|\Omega
|} \right]. \label{BTK}
\end{eqnarray}
Note that the terms $\propto {\cal C}_S,{\cal C}_1$ in Eq.
(\ref{g1}) are evaluated in the limit $T_{1,n},T_{2,n}\ll 1$ where
they only matter as compared to $g_1^{\rm BTK}(\varepsilon)$
provided $R_{1,2}\gg r$. The Cooperon term $\propto {\cal C}_1$
describes enhancement of Andreev conductance by electron
interference in diffusive N-metal \cite{VZK,HN,Z}, while the term
$\propto {\cal C}_S$ accounts for broadening of the density of
states in the superconductor. The spectral conductance
$g_2(\varepsilon)$ is given by Eq. (\ref{g1}) with interchanged
indices $1\leftrightarrow 2$.

Our next step is to solve the kinetic equation for the
distribution function $f_T$. In the limit $r/R_{1,2}\to 0$ this
solution is trivial: $f_T^S(\varepsilon)=0$ and
$f_T^{N_j}(\varepsilon,\bm{r}_j)=h(\varepsilon,V_j)
\equiv\left(\tanh[(\varepsilon+eV_j)/2T]-\tanh[(\varepsilon-eV_j)/2T]\right)/2$
($j=1,2$). In the first order in $r/R_{1,2}$ the function $f_T^S$
is determined from the diffusion equation
\begin{eqnarray}
\big(2\tilde\Omega - D\nabla^2\big) f_T=
\sum_{j=1,2}\frac{g_j(\varepsilon)h(\epsilon,V_j)}{2e^2N_SK(\varepsilon)}\delta(\bm{r}-\bm{r}_j),
\label{diffus}
\end{eqnarray}
where
$K(\varepsilon)=\theta(\Delta-|\varepsilon|)\Delta^2/\Omega^2-
\theta(|\varepsilon|-\Delta)\varepsilon^2/\Omega^2$, and
$\tilde\Omega=\theta(\Delta-|\varepsilon|)\Omega$ . Resolving Eq.
(\ref{diffus}) and substituting the result into Eq. (\ref{I1}), we
obtain
\begin{eqnarray}
I_{1}(V_1,V_2)=\int d\varepsilon \big[ g_{11}(\varepsilon)h(\varepsilon,V_1)-g_{12}(\varepsilon)h(\varepsilon,V_2)\big],
\end{eqnarray}
where
\begin{eqnarray}
g_{11}(\varepsilon) = g_1(\varepsilon) -\frac{{\cal
D}_S^{\bm{r}_1\bm{r}_1}(2i\tilde\Omega)}{2e^2N_S}\frac{
g_1^2(\varepsilon)}{K(\varepsilon)} -\frac{{\cal
D}_1^{\bm{r}_1\bm{r}_1}(0)}{2e^2N_S} g_1^2(\varepsilon),
\label{g11}
\\
g_{12}(\varepsilon) =g_{21}(\varepsilon)= \frac{{\cal
D}_S^{\bm{r}_1\bm{r}_2}(2i\tilde\Omega)}{2e^2N_S} \frac{
g_1(\varepsilon) g_2(\varepsilon)}{K(\varepsilon)},
\hspace{1cm}
\label{g12}
\end{eqnarray}
and similarly for the current $I_2$. The last two terms in the
local conductance $g_{11}(\varepsilon)$ (\ref{g11}) describe
partial conductance suppression respectively due to local charge
imbalance inside the superconductor and due to non-equilibrium
quasiparticles in the normal metal. At energies
$|\varepsilon|>\Delta$  Eq. (\ref{g12}) accounts for the effect of
{\it non-local} charge imbalance which yields non-zero
contribution to $g_{12}(\varepsilon)$ already in the lowest order
in $1/R_{1}R_{2}$. In contrast, at subgap energies this lowest
order contribution vanishes identically manifesting the well known
cancellation between EC and CAR terms \cite{FFH}. This
cancellation is lifted in higher orders in barrier transmissions
\cite{KZ06}. Accordingly, the full expression for
$g_{12}(\varepsilon)$ (\ref{g12}) does not vanish also for
$|\varepsilon|<\Delta$ and describes non-trivial interplay between
CAR and direct electron transfer in the presence of disorder.

Eq. (\ref{g12}) for the non-local spectral conductance -- together
with Eqs. (\ref{g1}), (\ref{BTK}) and (\ref{g11}) -- is the
central result of this work.  Note that this result is not
specific to particular geometry of Fig. 1 but applies for other
diffusive NSN structures as well.

\begin{figure}
\begin{tabular}{cc}
\includegraphics[width=4cm]{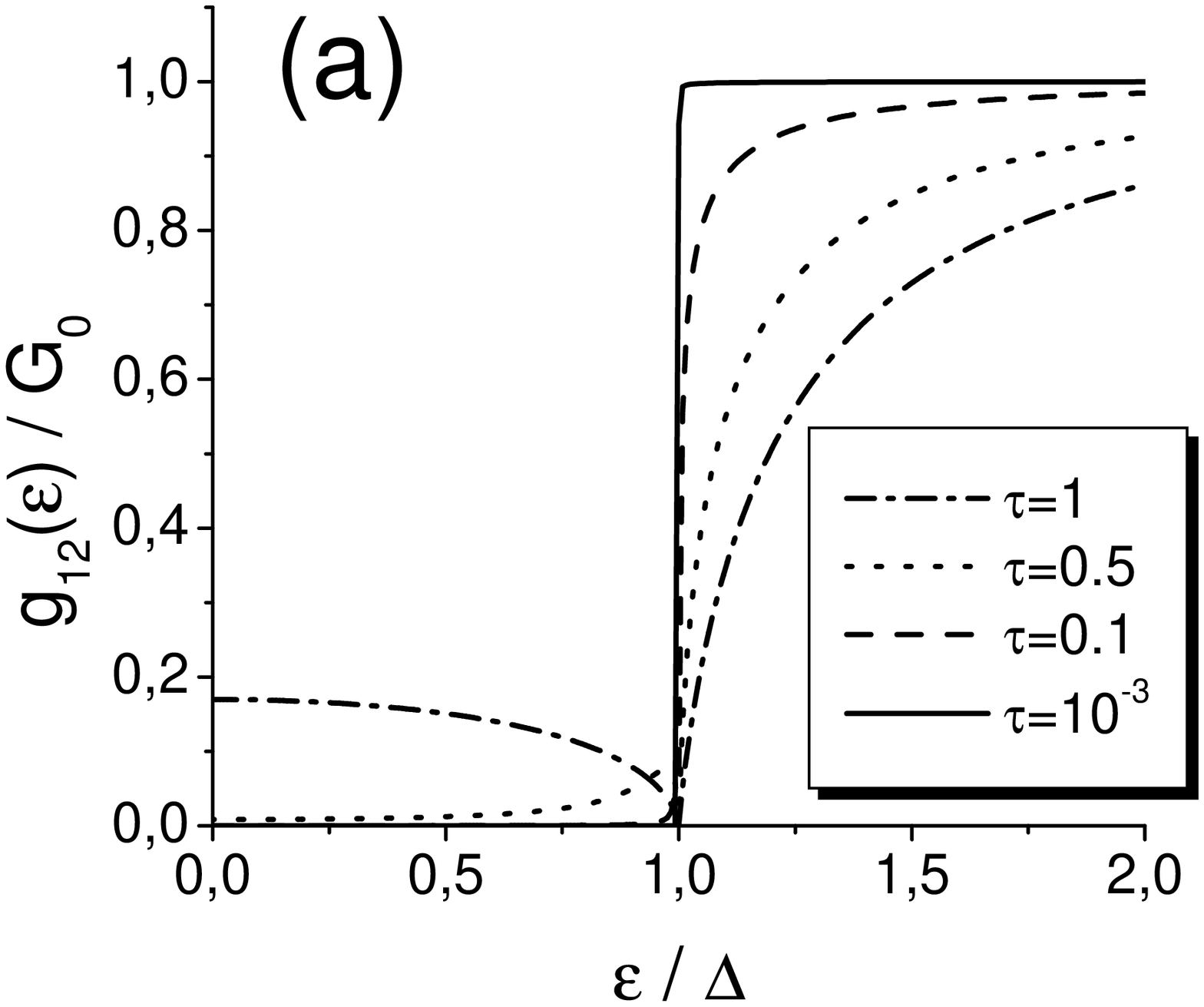} &
\includegraphics[width=4cm]{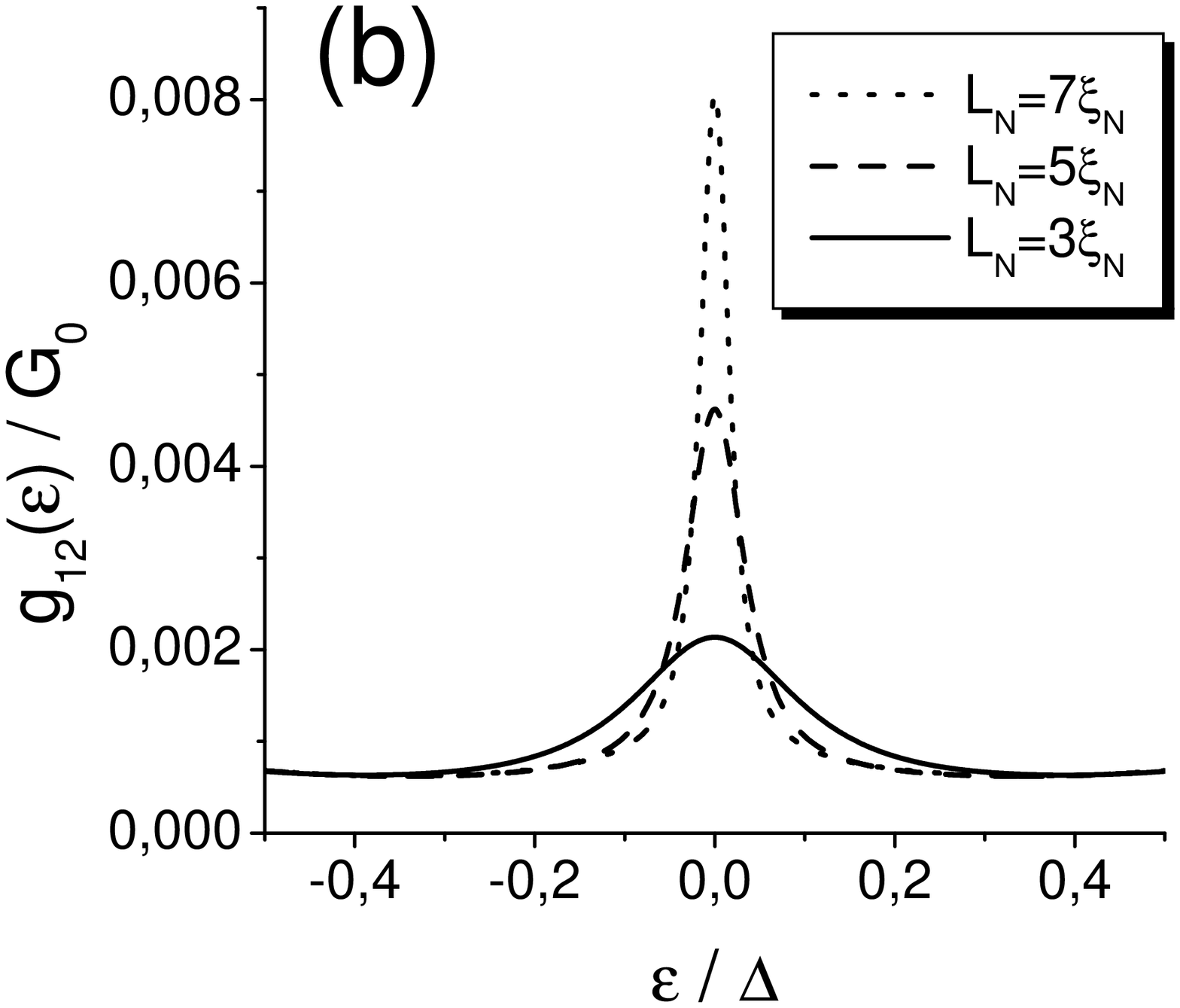}
\end{tabular}
\caption{Non-local spectral conductance $g_{12}(\varepsilon)$
(normalized by $G_0=r_{x_1}r_{L-x_2}/r_LR_1R_2$) for diffusive NSN
structures. We set $L=10\sqrt{2}\xi_S(0)$, $x_1=0.45L$, and
$x_2=0.55L$. (a) The case of two identical barriers with
resistances $R_1=R_2=\pi /e^2 N_{ch} \tau$ ($N_{ch}$ is the number
of channels and $\tau$ is the barrier transmission) and for
$r_{N_1}=r_{N_2}=0$. (b) The case of two tunnel barriers with
$T_{1,n},T_{2,n}\ll 1$ and for $L_1=L_2=L_N$, $r_{N_1}=r_{N_2}$,
$R_1=R_2$, $r_{N1}\xi_{N}/L_1R_1=0.0025$ and
$\sqrt{2}r_L\xi_S/LR_1=0.005$. }
\end{figure}

Here, however, we will only analyze the system with effectively
quasi-one-dimensional superconducting and normal wires, as shown
in Fig. 1. Assuming $x_2>x_1$ we obtain
\begin{eqnarray}
{\cal D}_S^{x_1x_2}(\omega) &=&
\frac{\sinh [k(L-x_2)]\sinh kx_1}{kS_SD_S\sinh (kL)},
\nonumber\\
{\cal C}_{j}^{x_{j}x_{j}}(\omega) &=&
\frac{\tanh\left(\sqrt{(-i\omega+1/\tau_{\varphi})/D_{j}}L_{j}\right)}{S_{j}D_{j}\sqrt{(-i\omega+1/\tau_{\varphi})/D_{j}}},
\label{DC1d}
\end{eqnarray}
where $j=1,2$. Here $S_{S,1,2}$ and $D_{S,1,2}$ are respectively
effective cross sections and diffusion coefficients of the
corresponding terminals and $k=\sqrt{(-i\omega+1/\tau_{Q^*})/D_S}$.
Substituting Eqs. (\ref{DC1d}) into (\ref{g11})-(\ref{g12}) we
arrive at the conductance matrix describing the system in Fig. 1.

{\it Zero-bias anomaly.} Let us first analyze the tunneling limit
$T_{1,n},T_{2,n} \ll 1$. In this case at subgap energies the term
$g_1^{\rm BTK}(\varepsilon)$ (\ref{BTK}) can be neglected and for
$E_{1,2}\equiv D_{1,2}/L_{1,2}^2\ll |\varepsilon|<\Delta $ we
obtain
\begin{eqnarray}
g_{11}(\varepsilon)&=&\frac{\Delta^2}{\Omega^2}
\left[\frac{r_{\xi_S}(\varepsilon)+r_{\xi_1}(\varepsilon)}{2R_1^2}
+\frac{r_{\xi_S}(\varepsilon)}{2R_1R_2}e^{-\frac{|x_2-x_1|}{\xi_S(\varepsilon)}}\right]
\label{1dg11}\\
g_{12}(\varepsilon)&=&\frac{\Omega^2}{2\Delta^2}
r_{\xi_S}(\varepsilon)g_{11}(\varepsilon)g_{22}(\varepsilon)
e^{-\frac{|x_2-x_1|}{\xi_S(\varepsilon)}} . \label{1dg12}
\end{eqnarray}
Here $r_{\xi_S}(\varepsilon)=r_L\xi_S(\varepsilon)/L$ and
$r_{\xi_{1,2}}(\varepsilon)=r_{N_{1,2}}\xi_{1,2}(\varepsilon)/L_{1,2}$
are Drude resistances of the segments of S- and N-metals with
respective lengths $\xi_S(\varepsilon)=\sqrt{D_S/2\Omega}$ and
$\xi_{1,2}(\varepsilon) =\sqrt{D_{1,2}/|\varepsilon |}$.

At small energies the local spectral conductance diverges as
$g_{11}(\varepsilon) \propto 1/\sqrt{\varepsilon}$ which is just
well known disorder-induced zero-bias anomaly \cite{VZK,HN,Z}. For
the non-local conductance (\ref{1dg12}) this divergence turns out
to be even stronger, $g_{12}(\varepsilon) \propto 1/\varepsilon$,
since quantum interference in {\it both} diffusive normal metals
simultaneously enhances non-local electron transport in our
system. Thus, we predict {\it a sharp low energy peak} in the
non-local conductance which occurs in the presence of disorder in
the N-terminals, see also Fig. 2. Accordingly, the differential
conductance $G_{12}(V_2,T)=-\partial I_1/\partial V_2$ increases
as $G_{12}\propto 1/{\rm max} (eV_2,T)$ with decreasing voltage
and temperature.

Eq. (\ref{1dg11}) applies down to $\varepsilon \sim E_1$ and for
even smaller energies $r_{\xi_1}(\varepsilon)$ should be
substituted by $2r_{N_1}$. Then for $r_{N_1}, r_{N_2} \gg
r_{\xi_S} \equiv r_{\xi_S}(0)$ we get
\begin{equation}
g_{12}(0)=G_{12}(0,0)=\frac{r_{\xi_S}r_{N_1}r_{N_2}}{2R_1^2R_2^2}
e^{-{|x_2-x_1|}/{\xi_S}}.
\end{equation}
We also note that in the case of strongly asymmetric barriers $R_2
\ll R_1$ the dominating contribution to $g_{12}$ scales as
$\propto 1/R_1R_2^3$ rather than $\propto 1/R_1^2R_2^2$.

Turning to the case of high barrier transmissions $T_{1,n},T_{2,n}
\lesssim 1$ we observe that in this case $g_1$ is dominated by
$g_1^{BTK}$ (\ref{BTK}) while other contributions can be
neglected. In particular, for fully open barriers at subgap
energies we obtain $g_{1,2}=g_{1,2}^{BTK}=2/R_{1,2}$ and, hence,
\begin{equation}
g_{12}(\varepsilon
)=\frac{\Omega^2}{\Delta^2}\frac{2r_{\xi_S(\varepsilon )}}{R_1R_2}
e^{-{|x_2-x_1|}/{\xi_S(\varepsilon)}},\quad |\varepsilon | <\Delta
.
\end{equation}

\begin{figure}
\begin{tabular}{cc}
\includegraphics[width=4cm]{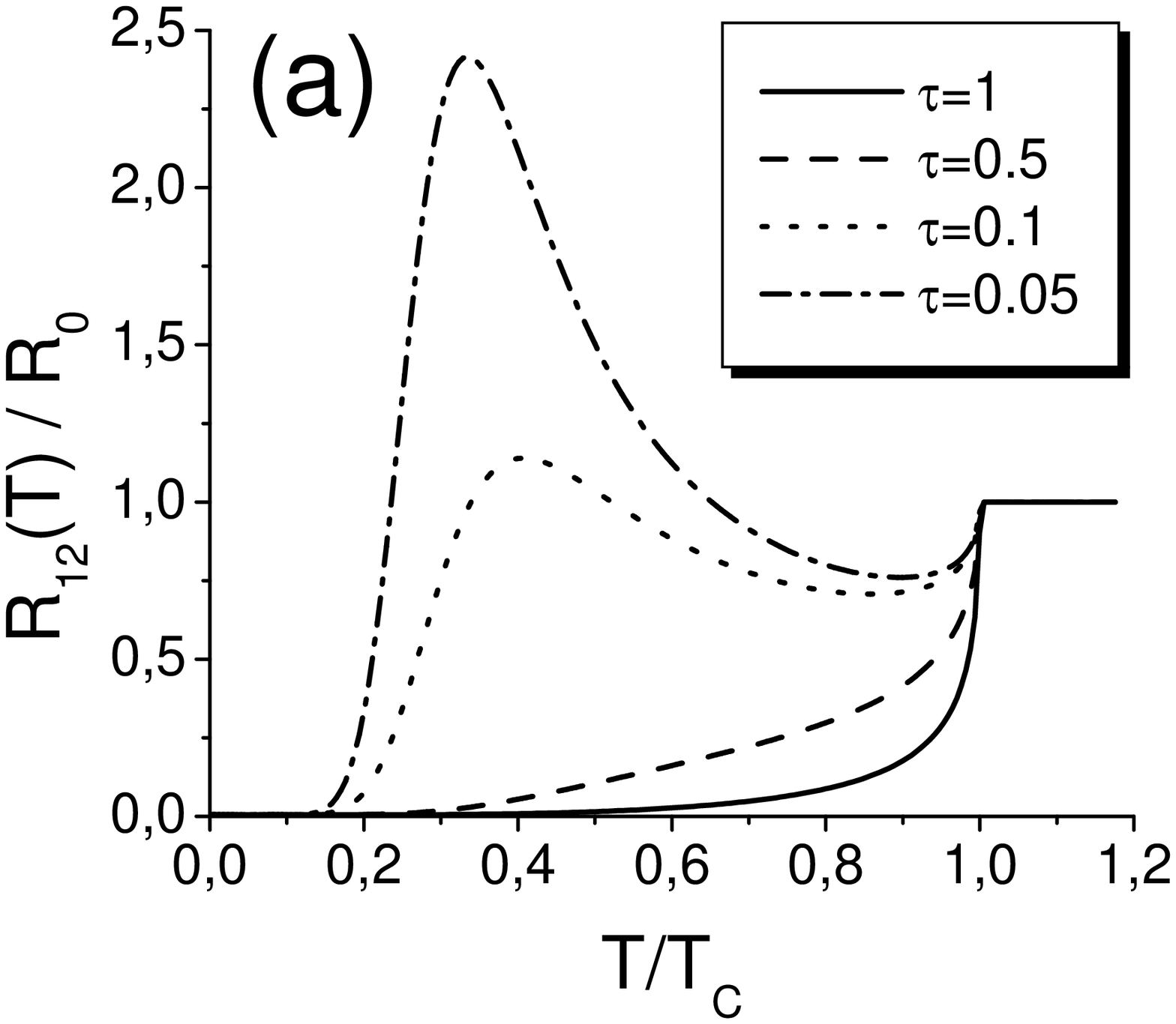} &
\includegraphics[width=4cm]{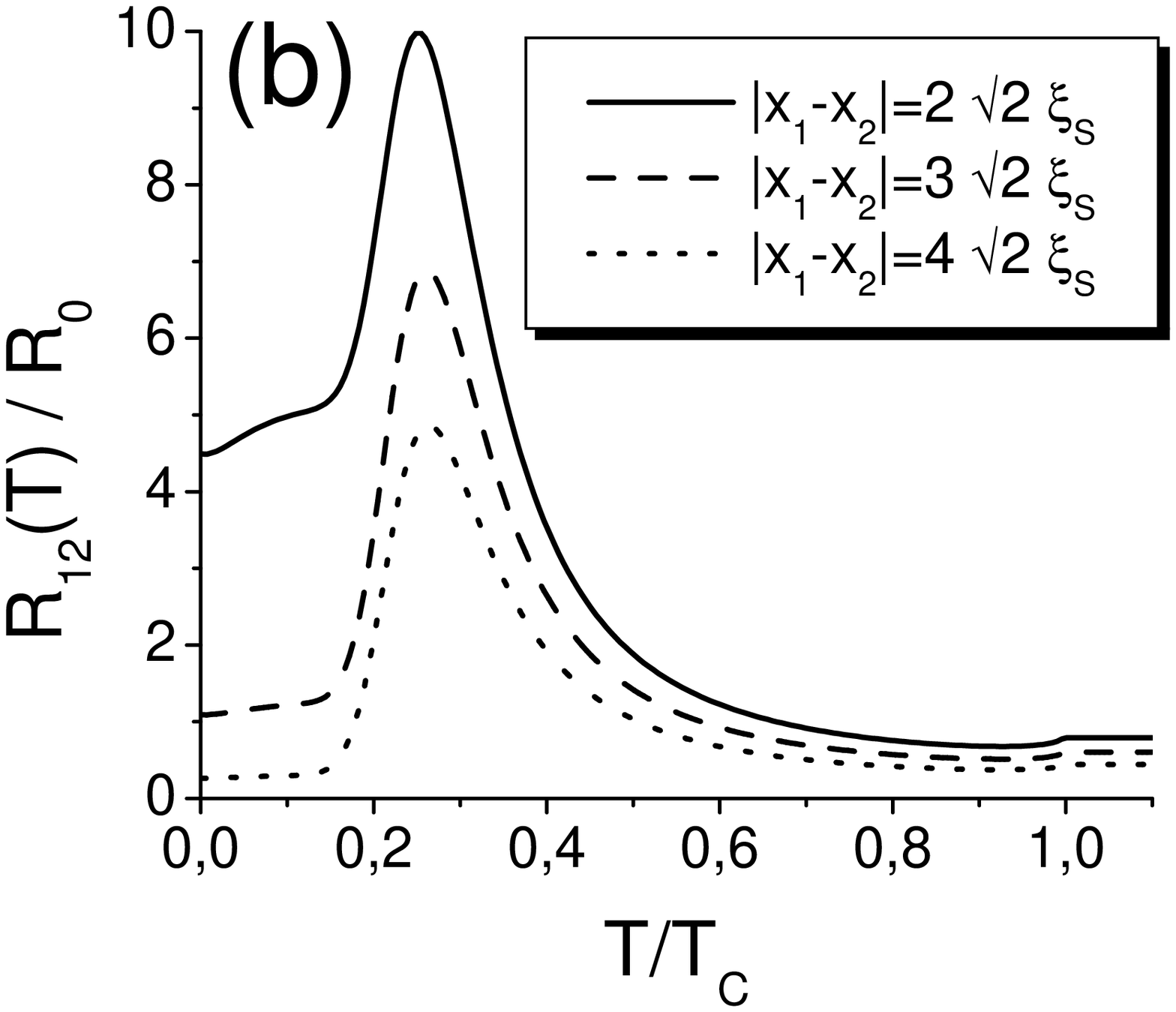}
\end{tabular}
\caption{ (a) Non-local resistance $R_{12}(T)$ (\ref{Rnl}), normalized by
$R_0=r_{x_1}r_{L-x_2}/r_L$,  at
different barrier transmissions $\tau$ and at $\tau_{Q^*},\tau_{\varphi}\to\infty$.
We chose $R_1=R_2=\pi /e^2
N_{ch} \tau=h/20e^2$ and set $L=20\sqrt{2}\xi_S(0)$,
$x_1=9\sqrt{2}\xi_S(0)$, $x_2=11\sqrt{2}\xi_S(0)$, $r_L=200$
$\Omega$ and $r_{N_1}=r_{N_2}=0$. (b) $R_{12}(T)/R_0$ at different
$|x_2-x_1|$. Other parameters are the same as in
Fig. 2b.}
\end{figure}

{\it Non-local resistance and charge imbalance peak.} Let us now
define non-local linear resistance
\begin{eqnarray}
R_{12}(T)=\frac{G_{12}(0,T)}{G_{11}(0,T)G_{22}(0,T)-G_{12}(0,T)G_{21}(0,T)}.
\label{Rnl}
\end{eqnarray}
Combining this equation with Eq. (\ref{g12}), at $T\ll \Delta$ we
arrive at a very simple and universal formula
\begin{eqnarray}
R_{12}= (r_{\xi_S}/2)\, e^{-{|x_2-x_1|}/{\xi_S}}. \label{R012}
\end{eqnarray}
It is remarkable that {\it independently of both barrier and
N-terminal parameters} the subgap non-local resistance is set only
by the normal state resistance $r_{\xi_S}$ of the superconducting
wire segment of length $\xi_S$ and by the distance between the
barriers measured in units of $\xi_S$. At low $T$ the dependence
$R_{1,2}\approx r_0\exp[-|x_2-x_1|/\xi_S(0)]$ was observed in
experiments \cite{Venkat} with $r_0\approx 0.56$ $\Omega$. For the
parameters \cite{Venkat} we estimate $r_0=r_{\xi_S}/2$ in the
range of one $\Omega$. A similarly good agreement is found between
Eq. (\ref{R012}) and experimental results \cite{Beckmann}.

The temperature dependence of $R_{12}(T)$ is depicted in Fig. 3.
In the tunneling limit it exhibits a well pronounced peak which
originates from the competition between charge imbalance and
Andreev reflection \cite{GZ07,KZ08}. The maximum value of the
non-local resistance $R_{12}$ is reached at $T^*\simeq 2\Delta/\ln
(R_1R_2/r^2_{\xi_S})$ and reads
\begin{equation}
R_{12}(T^*) \approx \frac{\alpha
r_{x_1}\sqrt{\frac{8T^*}{\pi\Delta}}\left(1-\frac{|x_2-x_1|}{\lambda}
\right) }
{\left(\sqrt{\frac{r_{\xi_S}+r_{\xi_1}(T^*)}{R_1}}+\sqrt{\frac{r_{\xi_S}+r_{\xi_2}(T^*)}{R_2}}\right)^2},
\end{equation}
where $\lambda =\alpha L$ and $\alpha =1-r_{x_1}/r_L$. Thus, the
peak resistance $R_{12}(T^*)$ decreases linearly with increasing
distance $|x_2-x_1|$ between the barriers. This behavior agrees
well with recent observations \cite{Venkat}. Furthermore, with the
parameters \cite{Venkat} we estimate $\lambda$ to be of order a
micron in agreement with experimental findings. Such values of
$\lambda$ appear lower than typical values of the charge imbalance
relaxation length $\Lambda_{Q^*}=\sqrt{D_S\tau_{Q^*}}$. The latter
length scale is expected to gain importance only for
$\Lambda_{Q^*}<\lambda$.

In summary, we developed a microscopic theory of non-local
electron transport in diffusive NSN systems which accounts for
non-trivial interplay between crossed Andreev reflection,
disorder, quantum interference and non-local charge imbalance. Our
results can be directly used for quantitative analysis of future
experiments.

This work was supported in part by RFBR grant 09-02-00886. D.S.G.
and M.S.K. also acknowledge support respectively from DFG Center
for Functional Nanostructures (CFN) and from the Dynasty
Foundation.

\end{document}